\documentclass[a4paper,twoside,11pt]{article}
\usepackage[bindingoffset=0.5cm,textheight=22.6cm,hdivide={2.9cm,*,2.9cm}, vdivide={*,22cm,*}]{geometry}
\usepackage{amsmath}
\usepackage{amsfonts}
\usepackage[dvips]{graphicx}
\usepackage[dvips,bookmarksnumbered=true,breaklinks=true]{hyperref}

\usepackage[square,comma,sort&compress]{natbib}

\newcommand{\uim}{UV/IR mixing}
\newcommand{\nc}{non-com\-mu\-ta\-tive}
\newcommand{\etal}{{\it et al.}}
\newcommand{\eqnref}[1]{Eqn.~(\ref{#1})}		
\newcommand{\secref}[1]{Section~\ref{#1}}		
\newcommand{\starco}[2]{\left[ #1\stackrel{\star}{,}#2\right] }		
\newcommand{\pa}{\partial}						
\newcommand{\ri}{{\rm i}}						
\newcommand{\re}{{\rm e}}						
\renewcommand{\k}{\tilde{k}}						
\newcommand{\p}{\tilde{p}}						
\newcommand{\Dt}{\widetilde{D}}						
\newcommand{\Act}{S}
\renewcommand{\th}{\theta}

\renewcommand{\l}{\lambda}
\newcommand{\m}{\mu}
\newcommand{\n}{\nu}

\renewcommand{\r}{\rho}
\newcommand{\s}{\sigma}

\renewcommand{\Xi}{\Xi}

\newcommand{\inv}[1]{\frac{1}{#1}}				

\newcommand{\intx}{\int\!\! {\rm d}^4x}						
\newcommand{\F}{\widetilde{F}}

\newcommand{\wsq}{\widetilde{\square}}
\newcommand{\ig}{{\rm i}g}

\title{\vspace{2cm}On the Problem of Renormalizability in Non-Commutative Gauge Field Models --- A Critical Review}
\author{Daniel N. Blaschke,
Erwin Kronberger, Arnold Rofner, Manfred Schweda,\\*[12pt] Ren\'e I.P. Sedmik and Michael Wohlgenannt}
\date{August 12, 2009}
\graphicspath{{./figures/}}
\begin{document}
\maketitle
\thispagestyle{empty}
\begin{center}
\renewcommand{\thefootnote}{\fnsymbol{footnote}}
\vspace{-0.3cm}
Institute for Theoretical Physics,
Vienna University of Technology\\
Wiedner Hauptstrasse 8-10, A-1040 Vienna (Austria)\\[0.3cm]
\ttfamily{E-mail: blaschke@hep.itp.tuwien.ac.at, kronberger@hep.itp.tuwien.ac.at, arofner@hep.itp.tuwien.ac.at, mschweda@tph.tuwien.ac.at, sedmik@hep.itp.tuwien.ac.at, miw@hep.itp.tuwien.ac.at}
\vspace{0.5cm}
\end{center}%
\begin{abstract}
When considering quantum field theories on {\nc} spaces one inevitably encounters the infamous {\uim} problem. So far, only very few renormalizable models exist and all of them describe {\nc} scalar field theories on four-dimensional Euclidean Groenewold-Moyal deformed space, also known as `$\th$-deformed space' $\mathbb{R}^4_\th$. In this work we discuss some major obstacles of constructing a renormalizable {\nc} gauge field model and sketch some possible ways out.
\end{abstract}

%
\section{Introduction}
Ever since the first {\nc} quantum field theory models were constructed, the greatest obstacle has been the infamous so-called {\uim} problem~\cite{Minwalla:1999}, where certain types of Feynman graphs, the non-planar graphs, exhibit new unrenormalizable IR singularities in exceptional momenta (see~\cite{Douglas:2001,Szabo:2001,Rivasseau:2007a} for a review). The situation improved dramatically when the first renormalizable scalar {\nc} model, the Grosse-Wulkenhaar model, was put forward~\cite{Grosse:2003,Grosse:2004b,Rivasseau:2005a}. Later, a second renormalizable scalar {\nc} quantum field theory was presented by Gurau \emph{et al.}~\cite{Gurau:2009} which, due to the way it is constructed, we would like to refer to as the `scalar $1/p^2$ model'. Both models have in common that they are formulated on Groenewold-Moyal deformed~\cite{Groenewold:1946,Moyal:1949} (also called $\th$-deformed) Euclidean space $\mathbb{R}^4_\th$ (rather than Minkowski), and that they 'mix' short and long distances  which 'damps' potential IR divergences. While translation invariance is broken explicitly in the Grosse-Wulkenhaar model by adding an oscillator-like term to the action, the scalar $1/p^2$ model avoids this problem through a non-local bilinear term of the form $\phi\star \frac{a}{\square}\phi$ for the quadratic one-loop IR divergence inherently generated by the phase factors of the non-planar part at one-loop level. On the other hand, the Grosse-Wulkenhaar model implements the so-called Langmann-Szabo duality~\cite{Szabo:2002} and kills the infamous Landau ghost~\cite{Grosse:2004a,Rivasseau:2006b}, whereas the scalar $1/p^2$ model does not. Nonetheless, both models have been proven to be renormalizable to all orders of perturbation theory.

Despite of almost a decade of work in this field two important steps have not been achieved yet:
\begin{itemize}
\item a good handling and efficient computation of Feynman diagrams on {\nc} Minkowski space-time and the construction of a candidate for a renormalizable scalar model\footnote{There have been claims, that the UV/IR mixing is not present in a Minkowskian {\nc} QFT if one considers proper Feynman rules taking into account a generalized notion of time ordering~\cite{Liao:2002,Fischer:2002,Bahns:2002,Denk:2003,Bahns:2003,Bahns:2004}. However, these conjectures still lack a rigorous proof.}, such as a Minkowskian version of the Grosse-Wulkenhaar or the scalar $1/p^2$ model, although for the latter a promising candidate has recently been put forward~\cite{Szabo:2008};
\item the construction of a renormalizable gauge model, or more precisely, a rigorous proof of renormalizability of one of the promising candidates~\cite{Blaschke:2007b,Grosse:2007,Wulkenhaar:2007,Vassilevich:2008,Blaschke:2008a,Blaschke:2009a,Blaschke:2009b}.
\end{itemize}
In this work we shall discuss the present status of {\nc} gauge theories and point out obstacles for renormalization on a very general basis, i.e. not particular to any specific model. We presume $\theta$-deformed $\mathbb{R}_\theta^4$ endowed with the non-local Groenewold-Moyal star product~\cite{Groenewold:1946,Moyal:1949}
\begin{align}
\left(\Phi_1\star\Phi_2\right)(x)&\equiv\Phi_1(x)\,\re^{\frac{\ri}{2}\overleftarrow{\partial}_\mu\th_{\mu\nu}\overrightarrow{\pa}_\nu}\Phi_2(x)\,,\qquad\text{where }\quad
\ri\th_{\m\n}\equiv\starco{x_\m}{x_\n}\,.
\label{eq:star_product}
\end{align}
In the simplest case, which we would like to consider here, the antisymmetric real matrix $\th_{\m\n}$ is constant and has mass dimension $-2$. In addition we would like to introduce the notation $\p_\m\equiv\theta_{\m\n}p_\n$.

We start by discussing the damping of IR singularities in {\nc} scalar QFTs, and consider the scalar $1/p^2$ model as an example in \secref{sec:scalar}. We then move on to gauge theories and the numerous unsolved problems accompanying them in Sections~\ref{sec:gauge} and~\ref{sec:locality} before, in \secref{sec:roadmap}, we finally try to find a roadmap for future strategies concerning the construction of renormalizable gauge models on $\th$-deformed spaces.

\section{Implications of UV/IR mixing}
\label{sec:uvir}
In order to find a way out of the UV/IR problems we have to fully understand the mechanisms leading to the mixing, and the implications originating from it. Obviously, the $\inv{\p^2}$ singularity is intimately tied to the Groenewold-Moyal product. More specifically, the deformation gives rise to phase factors of the type $\re^{\ri k_\m\th_{\m\n}p_\n}$,  with $k_\m$ being an internal momentum to be integrated out, and $p_\m$ being an external momentum. In the UV limit the rapid oscillation effectively eliminates UV divergences in (non-planar) loop integrals. However, this damping behaviour vanishes in the limit $p_\m\to0\;\forall\m$ or $\p_\m\to0\;\forall\m$, where the phase becomes unity. Naturally, in this limit the original divergence has to reappear, and in the case of a quadratic divergence is represented in the form $\inv{\p^2}$.

Historically, (an incomplete list of references is given by~\cite{Martin:2000,Grosse:2000,Liao:2001,Armoni:2000xr,Matusis:2000jf}) the IR divergences have been neglected in the discussion of renormalization. Instead, direct correspondences between known commutative results and the outcome of planar part calculations of the {\nc} counter parts have been sought. Soon afterwards, there appeared a series of publications~\cite{Minwalla:1999,VanRaamsdonk:2000,Szabo:2001,Schweda:2002b,Nakajima:2002} describing the, finally discovered, UV/IR mixing in all detail. However, it was not clear at this point how to apply renormalization in the presence of this new effect due to the following: On the one hand, the IR divergences are no `classical' singularities appearing in some ill-defined loop integrals requiring regularization, but divergences in the \emph{external} momentum. Therefore, the well known renormalization schemes from commutative quantum field theory cannot be applied straightforwardly. On the other hand, the standard choice for the renormalization conditions~\cite{Piguet:1995,Schweda-book:1998} cannot be taken due to the appearance of the $1/\p^2$ term in loop corrections. We will come back to these problems\footnote{One should also mention, that supersymmetry can in principle improve the situation by reducing the degrees of divergences, and hence the UV/IR mixing. (An incomplete list of references is given by~\cite{SheikhJabbari:1999iw, Rivelles:2000, Zanon:2000, Schweda:2002c,Ferrari:2004ex}.)} in \secref{sec:roadmap}, but for now, let us review another aspect of {\uim}: the problem of non-renormalizability, and damping as its cure.

%
\subsection{Damping in the scalar \texorpdfstring{$1/p^2$}{1/p**2} model}
\label{sec:scalar}
As has been discussed at the beginning of this section, models formulated on a deformed space, e.g. on Euclidean $\mathbb{R}^4_\th$, typically exhibit non-local infrared divergences due to {\uim} which lead to non-local counter terms. In the `na\"ive' version of {\nc} $\phi_\th^4$, which is generated by simply replacing ordinary products by their star-equivalents, one obtains the same propagator $G(k)=(k^2+m^2)^{-1}$ as in the commutative case. Computing a simple tadpole graph with $n$ non-planar insertions results in~\cite{Minwalla:1999,Micu:2000,Blaschke:2008b}
\begin{align}
 \raisebox{-10pt}{\includegraphics[scale=0.7, trim=10 5 8 5]{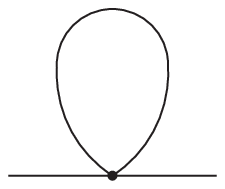}}\,\propto\inv{\tilde p^2}\;\,\Rightarrow\;\,\raisebox{-8pt}{\includegraphics[scale=0.7, trim=0 10 25 8]{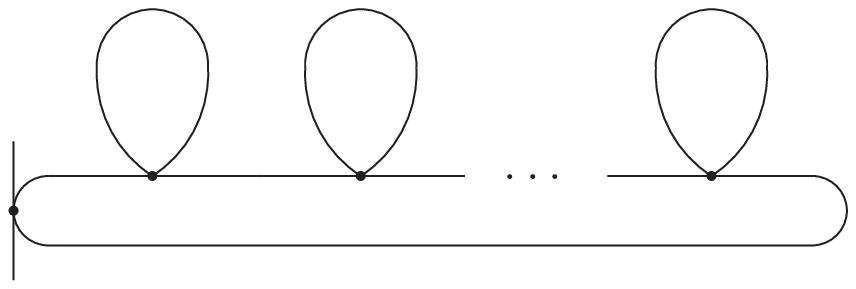}}\propto 
\int\!\! d^4p \inv{(\tilde p^2)^n}\inv{(p^2+m^2)^{n+1}}\,.
\label{eq:naive_tadpole}
\end{align}
Hence, the IR singularity grows order by order. This is due to the fact that there is no 
term present in the action of na\"ive $\phi_\th^4$ which damps the propagator. In the case of the scalar $1/p^2$ model of Gurau {\em et al.}~\cite{Gurau:2009}, for example, such an IR damping term is already included in the tree level action, which leads to the renormalizability of the model. This can be understood when considering their action
\begin{align}\label{eq:gur-mod}
\Act_{\text{Gur.}}&=\intx\left[\inv{2}\pa_\m\phi\star\pa_\m\phi+\inv{2}m^2\phi^{\star2}-\inv{2}\phi\star\frac{a^2}{\wsq}\phi+\frac{\l}{4!}\phi^{\star4}\right]\,,
\end{align}
whose bi-linear part leads to the propagator
\begin{align}\label{eq:gur-prop}
G^{\phi\phi}(k)&=\inv{k^2+m^2+\frac{a^2}{\k^2}}\,.
\end{align}
The divergence structure of the one-loop calculations is of course dominated by the region of large $k$, i.e. where the $a$-dependent part of the propagator is negligible, and for the self-energy one finds a quadratic UV divergence from the planar sector and a UV finite contribution from the non-planar sector. This latter contribution again behaves like $1/\p^2$ for small external momentum $p$, i.e. at $\p^2=0$ one has a quadratic IR divergence in the external momentum. However, the model \eqref{eq:gur-mod} already features an according (compensating) term in the tree-level action, and therefore the non-planar IR divergence merely leads to a renormalized parameter $a$~\cite{Tanasa:2008a,Blaschke:2008b}. Of course, it has to be assured that no additional divergences appear at higher loop order. In fact, recomputing \eqnref{eq:naive_tadpole} with the `damped' propagator \eqref{eq:gur-prop} (which behaves like $\k^2/a^2$ for small momentum $k$) yields an IR finite result independent of the number $n$ of insertions. A rigorous proof for this handwaving argumentation has been given up to all orders of perturbation theory by Gurau \etal~\cite{Gurau:2009} using Multiscale Analysis (MSA).

From the short discussion above we may learn three things:
\begin{itemize}
 \item Introduction of a $\theta$-deformed product into quantum field theories results in inherent non-locality. Therefore, the renormalization schemes known from commutative theory\footnote{In this respect we give a list of the most common technologies which, naturally, is far from being complete and shall just serve as a basic directory to start from: the Algebraic Renormalization procedure~\cite{Piguet:1995}, BPHZ subtraction procedure~\cite{Zimmermann:1969} and, equivalently, the bi-algebra based approach of Epstein and Glaser~\cite{Epstein:1973}.} that require locality, cannot be used directly.
 \item Nonetheless, it is possible to construct renormalizable models on {\nc} space --- at least in the scalar case as has been proven by using flow equations~\cite{Wilson:1973jj,Polchinski:1983gv} in a matrix base formulation~\cite{Grosse:2003,Grosse:2004b} and the MSA in $x$-space~\cite{Rivasseau:2005a,Gurau:2009}.
 \item The na\"ive approach of starting from a renormalizable commutative model and replacing all products with Groenewold-Moyal products does not lead to a renormalizable {\nc} theory. In fact, (by heuristic evidence) one is required to introduce additional non-local terms into the action in order to be able to absorb and damp the non-local IR divergences which are generated in loop calculations.
\end{itemize}

\subsection{Are non-commutative gauge theories feasible?}
\label{sec:gauge}
Let us consider the situation for gauge theories, and for simplicity start with $U(1)$ gauge fields on $\mathbb{R}^4_\th$. The gauge invariant Yang-Mills action endowed with stars, as would be the na{\"i}ve ansatz, is given by
\begin{align}\label{eq:naive-gauge-mod}
\Act_{\text{YM}\star}&=\intx\,\inv{4}F_{\m\n}\star F_{\m\n}\,,\qquad\text{with}\nonumber\\
F_{\m\n}&=\pa_\m A_\n-\pa_\n A_\m-\ig\starco{A_\m}{A_\n}\,.
\end{align}
Notice, that the star product \eqref{eq:star_product} modifies the initial $U(1)$ algebra in a way that 
it becomes non-Abelian\footnote{In $\th$-deformed $U(N)$ algebras all defining commutator relations are replaced by star-commutators $\starco{X^a(x)T^a}{Y^b(x)T^b}\equiv\left(X^a(x)T^a\star Y^b(x)T^b-Y^b(x)T^b\star X^a(x)T^a\right)$ where $X,Y$ are arbitrary functions on $\mathbb{R}_\th^4$, and $T$ are the generators of $U(N)$. Hence, even in the special case $N\to1$ these commutators do not vanish in contrast to the commutative case.}. We call the resulting algebra $U_\star(1)$. It has been shown~\cite{Terashima:2000,Matsubara:2000gr,Bars:2001,Chaichian:2001mu,Bonora:2000td} that only enveloping algebras, such as $U(N)$ or $O(N)$ and $USp(2N)$, survive the introduction of a deformed product (in the sense that commutators of algebra elements are again algebra elements), while e.g. $SU(N)$ does not. Despite this fact, star-commutators generally do not vanish due to property \eqref{eq:star_product}. Hence, any Moyal-deformed gauge theory is of the non-Abelian type.

As indicated above, the action \eqref{eq:naive-gauge-mod} does not lead to a renormalizable model no matter how the gauge fixing and Fadeev-Popov terms are chosen. 
The reason, as in the na\"ive scalar case, is that one finds a quadratic IR divergence in the non-planar part of the self-energy (which is independent of the gauge fixing~\cite{Blaschke:2005b,Hayakawa:1999,Hayakawa:1999b,Ruiz:2000}). This divergence increases as $\p^{-2n}$ with loop order $n$ (cf. \eqref{eq:naive_tadpole}). Due to BRST invariance it takes the form
\begin{align}\label{eq:IR-div}
\Pi^{\text{IR}}_{\m\n}&\propto g^2\frac{\p_\m\p_\n}{(\p^2)^2}\,,
\end{align}
i.e. $p_\m\Pi^{\text{IR}}_{\m\n}=0$ due to $p_\m\p_\m=p_\m\th_{\m\n}p_\n=0$, and no term exists in the tree level action to absorb this divergence.
Hence, in contrast to the scalar theory where renormalizability can be restored by adding a simple non-local term (see \secref{sec:scalar}), gauge theories contain an additional requirement for counter terms regarding the tensor structure. Indeed, the form of \eqnref{eq:IR-div} cannot simply be generated by contracting $F_{\m\n}$ with $\theta_{\m\n}$ \cite{Blaschke:2008a} but requires `fine tuning' of the action. Several generalizations of the mechanisms working in scalar theory to Yang-Mills type models have been presented~\cite{Blaschke:2007b,Grosse:2007,Wulkenhaar:2007,Blaschke:2008a,Blaschke:2009a,Blaschke:2009b}, but proofs of renormalizability are still missing, and it is not obvious that all problems can be solved. At least, up to now there is no model which features a suitable term to absorb the one-loop divergence \eqnref{eq:IR-div}.

It has recently been suggested~\cite{Vilar:2009} to conduct a proof of renormalizability by exploiting the symmetry content of the theory by means of Algebraic Renormalization (AR). This latter point is questionable due to inherent uncertainties which have to be clarified before such an attempt may succeed. See \secref{sec:locality} for a discussion of these problems.

Finally, one may state that the obstacles are in principle the same for scalar and gauge models. However, a straightforward generalization of the solutions~\cite{Rivasseau:2005a,Gurau:2009} for the scalar versions does not work in the latter case due to the additional demand for gauge invariance. In commutative theories, several techniques have been established over the years to conduct renormalization in the presence of symmetries~\cite{Piguet:1995,Schweda-book:1998} but application of these is prevented by an inherent property introduced by the deformation \eqref{eq:star_product}: non-locality.

\subsection{The curse of non-locality}
\label{sec:locality}
%
As indicated above, the introduction of the $\theta$-deformed product \eqref{eq:star_product} inevitably leads to non-locality since the product itself is non-local\footnote{We should mention, though, that attempts have been made to make the star product \emph{local} by introducing a bifermionic non-commutativity parameter~\cite{Vassilevich:2007, Vassilevich:2008}.}. This property is not compatible with some of the classical renormalization schemes. The reason is, that the premise of locality is required to avoid the insertion of artificial operators of negative mass dimension, which in turn spoil renormalizability. Of course, this is exactly the same problem we are facing now in {\nc} QFT and the question arises if renormalizability exists in the presence of non-locality. An answer to this problem has been partly given, but let us first review the problems in more detail in order to understand the background.

The BPHZ scheme, and also other subtraction schemes, relies on locality. In this method, divergent loop integrals are rendered finite by subtracting the first \(n\) terms of its Taylor expansion, where \(n\) corresponds to the order of the divergence according to na{\"i}ve power counting. By absorbing the divergences in the parameters of the theory one obtains renormalized quantities. Obviously, this procedure is only well-defined if the convergence radius of the series does not vanish\footnote{This becomes clear when considering the simple non-local function $1/p^2$, which cannot be Taylor expanded around $p=0$.}, i.e. for local integrands. However, a generalized subtraction scheme for integrands, where the non-locality only stems from the star-product, might be feasible.

Another quite successful scheme is Algebraic Renormalization (AR) (see~\cite{Piguet:1995} for a review of the topic). It relies on the so-called Quantum Action Principle (QAP)~\cite{Lowenstein:1971,Lowenstein:1971b,DAttanasio:1996jd} which, in turn, is based on the assumption of locality. Therefore, unfortunately, the QAP does not exist in its usual form on {\nc} spaces. However, even assuming that proofs can be found, such that a QAP exists in Moyal-deformed theories, one has to show that the symmetry content of the theory at tree level is stable under quantum corrections, i.e. that the theory is free from anomalies. This latter point involves the computation of the cohomology~\cite{Becchi:1988,Dixon:1991,Barnich:2000zw} $H(s)=\mathop{\rm Ker}s/\mathop{\rm Im}s$ (see also \cite{Brandhuber:1994uf} for an exemplary application of this concept in commutative theories and further references) of the nilpotent BRST operator $s$. One has to show triviality of the respective cohomology group for ghost number 1 local functionals (i.e. anomalies), which again requires locality in all steps~\cite{Barnich:2000zw}. In this respect, some efforts for a generalization to {\nc} spaces have been made. For example, the notion of BRST cohomology and the Chern character has been introduced in~\cite{Perrot:1999} using Connes' notation of spectral triples~\cite{Connes:1994,Connes:2006a}. Another contribution has been the generalization of the descent equations describing Yang-Mills anomalies to {\nc} spaces~\cite{Langmann:1995}. It has also been shown~\cite{Martin:2000} that the symmetry content compatible with the QAP can be established for {\nc} $U_\star(N)$ theories and is invariant under an explicit one-loop UV renormalization.

For the actual AR to be applicable to {\nc} spaces two things have to be shown. First, the computation of the cohomology class has to be worked out rigorously for the ghost number 0 functionals $\mathcal{F}$, representing the most general quantum level action, to fulfill $s\mathcal{F}=0$. In addition, a proof that the triviality of the cohomology is sufficient to guarantee renormalizability in the presence of non-locality is missing as well.

The second point is more involved and of a more general nature, as it applies to \emph{all} {\nc} theories: It concerns the appearance of dimensionless operator insertions in the action. A parameter of non-commutativity $\th$ with mass dimension $-2$ allows to freely add composite field operators\footnote{Note that this also occurs in scalar field theories. For example, the non-local term $\square^{-1}\phi^2$ could be inserted into the tree level action to arbitrary power.} of zero mass dimension, such as $D^2\Dt^2$ or $\F^2$, to the action, where $\Dt_\m = D_\n\th_{\m\n}$ is a contracted covariant derivative and $\F=F_{\m\n}\th_{\m\n}$ is a field strength. Being invariant under all symmetries appearing in the QAP (and gauge transformations in general), there is no constraint or theorem preventing insertions of arbitrary powers of these operators both at tree level or as quantum corrections\footnote{work in progress}. 
This is the reason why the sufficiency of a trivial cohomology class for renormalizability has been questioned above in \secref{sec:gauge}. We would further like to point out that, due to this problem, standard `top down' renormalization schemes cannot work, as they start from the set of all possible counter terms, and restrict them by applying constraints. Since this set is {\it {a} priori} \emph{infinite} in the presence of invariant dimensionless insertions, the attempt to achieve a finite number of counter terms will fail, independent of the cohomology.

%
\section{A roadmap to {\nc} gauge theories}\label{sec:roadmap}

Based on the findings above in \secref{sec:uvir}, we may now focus on possible solutions. The intention is to point out approaches which have a chance to enable renormalization on {\nc} spaces.

At first, bearing in mind the problem of the in principle infinite set of invariant counter terms, one is led to the insight that a classical `top down' scheme, attempting to restrict this infinite set by constraints generated by symmetries, will not succeed. Instead, the conjecture must be to work 'bottom up', i.e. to start from the tree level action and to successively find an estimation for all possible loop results, as it is achieved by the \emph{MSA}. However, the latter explicitly breaks gauge invariance. In Ref.~\cite{Rivasseau:1991ub}, attempts to treat pure Yang-Mills models within this renormalization approach are described. The authors tried to construct Schwinger functions of the field in Feynman or Landau gauge with an IR cut-off. However, their approach was unsuccessful: On the one hand the functional integrals they obtained lacked sufficient positivity, and on the other hand the related Gribov problem~\cite{Gribov:1978} was not solved. 
Nonetheless we may suggest to use the MSA and try to handle the Gribov problem using a soft breaking mechanism~\cite{Baulieu:2009,Blaschke:2009a} similar to the one present in the Gribov-Zwanziger action~\cite{Zwanziger:1989,Zwanziger:1993,Dudal:2008}, and see if renormalizability can in principle be achieved. 

A further possibility would be to conduct the \emph{Polchinski} approach which has successfully been applied to the Grosse-Wulkenhaar model \cite{Grosse:2003,Grosse:2004b}. The case of (commutative) spontaneously broken $SU(2)$ Yang-Mills theory has been discussed by C.~Kopper and V.~F.~M\"uller \cite{Kopper:2009zw}. Their starting point was the classical BRST invariant action including all (i.e. a finite set of) counter terms satisfying certain symmetry constraints. Since the regularization (which is required in the Polchinski approach) breaks the local gauge symmetry explicitly, the counter terms are only required to be invariant under a global $SO(3)$ isosymmetry. The authors showed that this ansatz solves the flow equations to all orders by induction. In the case of {\nc} gauge theories the set of all possible counter terms is infinite, but one could choose a restricted, finite set of counter terms instead. Renormalizability would be established, if it could be shown that this finite set solves the flow equations.

Another feasible path is to re-establish the foundations for \emph{AR} in the non-local case. As we have already mentioned, the classification of anomalies by computation of the cohomology class $H^{(1)}$ of the BRST operator for general functionals (i.e. counterterms) with ghost number 1 has already been achieved \cite{Perrot:1999} but the proof for ghost number 0, (i.e. the action) is missing. In addition it has to be assured in a rigorous way that trivial cohomology alone is sufficient to prove the absence of anomalies, i.e. renormalizability. And if this turns out not to be true, one has to find out which additional requirements are necessary.
Finally, and this is the most important point, we have to find constraints to limit the appearance of insertions of massless operators into the action. In this context also the issue of field redefinitions might be important and maybe some classes of insertions can be rewritten as such redefinitions (cf. \cite{Bichl:2001cq} in the context of {\nc} $U_\star(1)$ gauge theory with Seiberg-Witten maps).

\section{Conclusion}\label{sec:conclusion}
%
We have reviewed the current status of renormalization of {\nc} quantum field models. The effect of {\uim} gives rise to the well-known quadratic IR divergence in the external momentum $p$. In gauge theories this singularity is endowed with a new type of (transversal) tensor structure $p_\r\theta_{\r\m}p_\s\theta_{\s\n}\equiv\p_\m\p_\n$ which cannot be absorbed in a straightforward manner. For scalar models it has been shown that this can be done by adding a corresponding non-local term into the tree level action. This leads to renormalizability because the insertion alters the propagator in a way that it `damps' in the IR limit. In this respect, mainly two approaches have been followed: The Grosse-Wulkenhaar model featuring an oscillator-like term~\cite{Grosse:2003,Grosse:2004b}, and the $1/p^2$ model by Gurau {\etal}~\cite{Gurau:2009}. Both have been proven to be renormalizable up to all orders using `bottom up' schemes, such as the so-called Multiscale Analysis and the Polchinski approach. From the point of view of renormalization the inherent non-locality introduced by the $\theta$-deformed Groenewold-Moyal product turns out to be a great obstacle since almost all `classical' procedures rely on the presumption of locality. It is well known that omitting the latter requirement generally leads to non-renormalizability since it allows for the insertion of arbitrarily high powers of massless operators into the action. This is exactly the problem one is facing in {\nc} theories. 
In addition, in gauge models, the mentioned insertions are completely invariant under any symmetry compatible with the well known Quantum Action Principle, BRST, or gauge transformations. At present, there is no criterion to rule out these terms, and the set of all possible counter terms, which is the starting point for `classical' renormalization schemes such as the Algebraic Renormalization programme, is \emph{a priori} infinite.

In order to find a way out of this misery, and towards renormalizability of {\nc} gauge models, we have made several suggestions. The first one is to use schemes, such as the Polchinski approach or Multiscale Analysis, which both should in principle work out for gauge theories. However, there are indications that, in addition to the standard gauge fixing, a soft breaking mechanism is required in order to avoid the Gribov problem violating positivity of functional integrals. This point will have to be studied more thoroughly before the mist clears.

Another approach is to rigorously prove that trivial cohomology automatically induces absence of anomalies, and renormalizability, even in the presence of a deformed product. Since the latter point is rather questionable one will have to find an additional criterion to restore validity of the Algebraic Renormalization procedure, or find some proper modification. Finally, it will be of great importance to investigate possible conditions to restrict the appearance of arbitrary powers of massless operator insertions, which affect \emph{all} {\nc} models on the market. Maybe, things also simplify a bit and hints can be obtained more easily if one considers two dimensional models for a start. 

\subsection*{Acknowledgements}
This work was supported by the `Fonds zur F\"orderung der Wissenschaftlichen Forschung' (FWF) under contract P20507-N16.
The authors are indebted to P.~Aschieri, L.~Castellani, P.~A.~Grassi, H.~Grosse, J.~Madore, A.~Much,  H.~Steinacker and F.~Vignes-Tourneret for enlightening discussions on the topic.


\end{document}